\documentclass{amsart}
\usepackage{amssymb}
\usepackage{fullpage}
\usepackage{graphicx}
\input pictex.sty
\newtheorem{thm}{Theorem}[section]

\newtheorem{lem}[thm]{Lemma}
\newtheorem{prop}[thm]{Proposition}
\theoremstyle{definition}

\theoremstyle{remark}

\numberwithin{equation}{section}

\newcommand{\cb}{\mathbb{C}}

\newcommand{\lam}{\lambda}

\newcommand{\al}{\alpha}

\begin{document}

\title{A Quick Derivation of the Loop Equations for Random Matrices}%
\author{N. M. Ercolani}%
\address{Dept. of Math, Univ. of Arizona, 520-621-2713, FAX: 520-626-5186}%
\email{ercolani@math.arizona.edu}%
\author{K. D. T-R McLaughlin}%
\address{Dept. of Math., Univ. of Arizona}%
\email{mcl@math.arizona.edu}

\thanks{K. D. T-R McLaughlin was supported in part by NSF
grants DMS-0451495 and DMS-0200749, as well as a NATO
Collaborative Linkage Grant "Orthogonal Polynomials: Theory,
Applications, and Generalizations" Ref no. PST.CLG.979738. N. M.
Ercolani was supported in part by NSF grants DMS-0073087.}


\begin{abstract}
The "loop equations" of random matrix theory are a hierarchy of
equations born of attempts to obtain explicit formulae  for
generating functions of map enumeration problems. These equations,
originating in the physics of 2-dimensional quantum gravity, have
lacked mathematical justification.  The goal of this paper is to
provide a complete and short proof, relying on a recently
established complete asymptotic expansion for the random matrix
theory partition function.
\end{abstract}
\maketitle

\section{Background and Preliminaries}

The study of the Unitary Ensembles (UE) of random matrices
\cite{Mehta}, begins with a family of probability measures on the
space of $N \times N$ Hermitian matrices. The measures are of the
form

$$
d\mu_\textbf{t} = \frac{1}{\widetilde{Z}_N}\exp\left\{-N \mbox{ Tr
} [V_\textbf{t}(M)]\right\} dM,
$$
where the function $V_\textbf{t}$ is a scalar function, referred
to as the potential of the external field, or simply the "external
field" for short.  Typically it is taken to be a polynomial, and
written as follows:

$$
V_\textbf{t} = V(\lambda; \ t_{1}, \ldots, t_{\upsilon})
 = \frac{1}{2} \lambda^{2} + \sum_{j=1}^{\upsilon} t_{j}\lambda^{j}
$$
where the parameters $\{t_{1},\ldots, t_{\upsilon}\}$ are assumed
to be such that the integral converges.  For example, one may
suppose that $\upsilon$ is even, and $t_{\upsilon}>0$. The
\emph{partition function} $\widetilde{Z}_N$, is the  normalization
factor which makes the UE measures be probability measures.

Expectations of conjugation invariant matrix random variables with
respect to these measures can be reduced, via the Weyl integration
formula, to an integration against a symmetric density over the
eigenvalues which has a form proportional to (\ref{I.001}), below:

\begin{eqnarray} \label{I.001}
\exp{ \left\{ -N^{2}\left[\frac{1}{N} \sum_{j=1}^{N}
V(\lambda_{j}; \ t_{1}, \ldots, t_{\upsilon})  - \frac{1}{N^{2}}
\sum_{j\neq \ell} \log{| \lambda_{j} - \lambda_{\ell} | } \right]
\right\} } d^{N} \lambda.
\end{eqnarray}
These latter multiple integrals can be more compactly expressed in
terms of kernels constructed from polynomials
$\left\{p_\ell(\lambda)\right\}$ orthogonal with respect to the
exponential weight $e^{-N V_\textbf{t}(\lambda)}$ \cite{Mehta}.
For instance, the fundamental matrix moments
$\textbf{E}\left(\mbox{Tr} M^j\right)$, where $\textbf{E}$ denotes
expectation with respect the measure $d\mu_{\bf t}$, are expressed
as
\begin{eqnarray} \label{matmoments}
  \textbf{E}\left(\mbox{Tr} M^j\right) &=& N\int_{-\infty}^\infty
  \lambda^j \rho^{(1)}_N(\lambda) d\lambda
\end{eqnarray}
where $\rho^{(1)}_N(\lambda)$ denotes the so-called
\emph{one-point function}

\begin{eqnarray} \label{one-point}
 \nonumber \rho^{(1)}_N (x) &=& \frac{d}{dx} \mathbb{E}\left(\frac{1}{N} \# \left\{j: \lambda_j \in \left(-\infty,
  x\right)\right\}\right)\\
  &=& \frac{1}{N} K_N(x,x)\\
 \nonumber \mbox{with the kernel} \,\,\, K_N(x,y) &=& e^{-\frac{N}{2}\left(V_\textbf{t}(x) +
  V_\textbf{t}(y)\right)}\sum_{\ell=0}^\infty p_\ell(x) p_\ell(y).
\end{eqnarray}
The symbol $\mathbb{E}$ denotes expectation with respect to the
normalization of the measure (\ref{I.001}) which is given by
dividing this family of measures by the corresponding family of
integrals:

\begin{eqnarray}\label{partition}
& & Z_{N}(t_{1},t_{2}, \ldots, t_{\upsilon}) =\\
\nonumber & & \int \cdots \int \exp{ \left\{
-N^{2}\left[\frac{1}{N} \sum_{j=1}^{N} V(\lambda_{j}; \ t_{1},
\ldots, t_{\upsilon})  - \frac{1}{N^{2}} \sum_{j\neq \ell} \log{|
\lambda_{j} - \lambda_{\ell} | } \right] \right\} } d^{N} \lambda.
\end{eqnarray}
We will sometimes refer to the following set of ${\bf{t}} =
(t_{1}, \ldots, t_{\upsilon})$ for which (\ref{partition})
converges. For any given $T>0$ and $\gamma
0$, define
\begin{equation*}
\mathbb{T}(T,\gamma) = \{ {\bf{t}} \in \mathbb{R}^{\upsilon}: \
|{\bf{t}} | \le T, \ t_{\upsilon} > \gamma \sum_{j=1}^{\upsilon-1}
|t_{j}|\}.
\end{equation*}

\bigskip

The leading order behavior of $Z_{N}(t_{1},t_{2}, \ldots,
t_{\upsilon})$ is rather classical, and is known for a very wide
class of external fields $V$ (see, for example, \cite{Johansson}).
We will require the following result.

\begin{thm}\label{thm:lead}
There is $T > 0$ and $\gamma > 0$ so that for all $\bf{t} \in
\mathbb{T}(T,\gamma)$, the following holds true:
\begin{enumerate}
\item
\begin{eqnarray}
\label{eq:lead01} && \lim_{N \to \infty} \frac{1}{N^{2}}
\log\{Z_{N}(t_{1},t_{2}, \ldots, t_{\upsilon}) \} = -I (t_{1}, \ldots, t_{\upsilon})
\end{eqnarray}
where
\begin{eqnarray}
\label{eq:lead02} I (t_{1}, \ldots, t_{\upsilon}) = &&
\inf_{\text{Borel measures  }\mu ,\mu\geq 0, \int d\mu =1} \left[
\int V(\lam ) d\mu (\lam ) -\int\int \log | \lam -\mu
|\, d\mu (\lam )\, d\mu (\eta )\right] .
\end{eqnarray}
\item There is a unique measure $\mu_V$ which achieves the infimum
defined on the right hand side of (\ref{eq:lead02}). This measure
is absolutely continuous with respect to Lebesgue measure, and
  \begin{align*}
d\mu_{V} &= \psi \, d\lam ,\\
\psi (\lam ) &= \frac{1}{2 \pi} \chi_{(\al ,\beta )} (\lam )
\sqrt{(\lam -\al )(\beta -\lam )}\, h(\lam ),
  \end{align*}
where $h(\lam )$ is a polynomial of degree $\upsilon -2$, which is
strictly positive on the interval $[\alpha,\beta]$ (recall that
the external field $V$ is a polynomial of degree $\upsilon$). The
polynomial $h$ is defined by
  \[
h(z) = \frac{1}{2\pi i }\, \oint\, \frac{V'(s)}{\sqrt{(s-\al )}
\sqrt{(s-\beta )}} \, \frac{ds}{s-z}
  \]
where the integral is taken on a circle containing $(\al ,\beta )$
and $z$ in the interior, oriented counter-clockwise.

\item There exists a constant $l$, depending on $V$ such that the
following variational equations are satisfied by $\mu_V$:
\begin{eqnarray}
\nonumber  \int 2 \log|\lambda - \eta|^{-1} d\mu_V(\eta) + V(\lambda)
&\geq& l \,\,\, \mbox{for}\,\,\, \lambda \in \mathbf{R}\backslash\mbox{supp}(\mu_V) \\
  \int 2 \log|\lambda - \eta|^{-1} d\mu_V(\eta) + V(\lambda)
  &=& l \,\,\, \mbox{for}\,\,\, \lambda \in \mbox{supp}(\mu_V). \label{Var}
\end{eqnarray}

\item The endpoints $\al$ and $\beta$ are determined by the
equations
  \begin{align*}
\int^{\beta}_{\al}\, \frac{V'(s)}{\sqrt{(s-\al )(\beta -s)}}\, ds &= 0\\
 \int^{\beta}_{\al}\, \frac{sV'(s)}{ \sqrt{(s-\al )(\beta -s)} }\, ds &=
2\pi .
  \end{align*}

\item The endpoints $\al ({\bf t}\, )$ and $\beta ({\bf t}\,
)$ are actually analytic functions of ${\bf t}$, which
possess smooth extensions to the closure of $\{ {\bf t}: {\bf t} \in
\mathbb{T}(T,\gamma) \}$.  They also satisfy
$-\alpha({\bf 0}) = \beta({\bf 0}) = 2$. In addition, the
coefficients of the polynomial $h(\lam )$ are also analytic
functions of ${\bf t}$, with smooth extensions to the
closure of $\{ {\bf t}: {\bf t} \in
\mathbb{T}(T,\gamma) \}$, with
  \[
h(\lam , {\bf t}={\bf 0}) = 1.
  \]
\end{enumerate}
\end{thm}
\bigskip

{\bf Remark}  The variational problem appearing in
(\ref{eq:lead02}) is a fundamental component in the theory of
random matrices, as well as integrable systems and approximation
theory.  It is well known, (see, for example, \cite{SaffTotik}),
that under general assumptions on $V$, the infimum is achieved at
a unique measure $\mu_{V}$, called the equilibrium measure. For
external fields $V$ that are analytic in a neighborhood of the
real axis, and with sufficient growth at $\infty$, the equilibrium
measure is supported on finitely many intervals, with density that
is analytic on the interior of each interval, behaving at worst
like a square root at each endpoint, (see \cite{DKM} and
\cite{DKMVZ3}). $\Box$
\medskip

\medskip

{\bf Remark}  For a proof of (\ref{eq:lead01}), we refer the
reader to \cite{Johansson}, however this result is commonly known
in the approximation theory literature. $\Box$
\medskip

{\bf Remark} It will prove useful to adapt the following
alternative presentation for the function $\psi$:
\begin{eqnarray} \label{psi}
\psi(\lambda) = \frac{1}{2 \pi i} R_{+}(\lambda) h(\lambda), \
\lambda \in (\alpha, \beta),
\end{eqnarray}
where the function $R(\lambda)$ is defined via $R(\lambda)^{2} = (
\lambda - \alpha) ( \lambda - \beta)$, with $R(\lambda)$ analytic
in $\cb \setminus [\alpha, \beta]$, and normalized so that $R(
\lambda) \sim \lambda$ as $\lambda \to \infty$.  The subscript
$\pm$ in $R_{\pm}(\lambda)$ denotes the boundary value obtained
from the upper (lower) half plane. $\Box$

\bigskip

The goal of this paper is to provide a rigorous justification for
the ``loop equations'' which originated in the physics of
2-dimensional quantum gravity (see, for example, the survey
\cite{Difrancesco95} and the references contained therein).  More
precisely, this entails
\begin{itemize}
\item Proving that the quantity $\int_{-\infty}^{\infty}
  \frac{\rho_{N}^{(1)}(x)}{x-z} dz$ possesses a complete asymptotic
  expansion in even powers of $N$ (see  Theorem \ref{1PTASSNew}, and
  (\ref{Pgexp})).
\item Establishing that the coefficients $P_{g}(z)$ in the
asymptotic
  expansion (\ref{Pgexp})  satisfy the hierarchy of nonlocal equations
  (\ref{LoopEQNFIN}), which are the loop equations.
\end{itemize}
Once the coefficients $P_{g}(z)$ are known to exist as analytic
functions of $z$ and the times ${\bf t}$, they may be interpreted
as generating functions for a collection of graphical enumeration
problems for labelled maps, counted according to vertex valences
and the genus of the underlying Riemann surface into which the
maps are embedded.  (See \cite{BIZ}, and also \cite{EM}.)  Because
of the combinatorial connection and its use in 2-dimensional
quantum gravity, obtaining explicit formulae has been a
fundamental goal within the physics community of quantum gravity.
The loop equations arose as a means to obtain explicit information
(and possibly explicit formulae) for these coefficients, although
without mathematical justification.

\medskip

In Section \ref{loop} we will need to consider the Cauchy
transform of the equilibrium measure:
\begin{eqnarray*}
  F(z) &=& \int_{-\infty}^\infty \frac{\psi(\lambda)}{z - \lambda}
  d\lambda, \,\,\, z \in \mathbb{C}\slash \mathbb{R}.
\end{eqnarray*}
It follows from differentiating the variational equations, Theorem
\ref{thm:lead}(3), that $F(z)$ solves the scalar Riemann-Hilbert
problem
\begin{eqnarray*}
  F_+(s) + F_-(s) &=& V^\prime(s),\,\,\, s \in [\alpha, \beta] \\
  F_+(s) - F_-(s)&=& 0, \,\,\,\,\,\,\,\,\,\,\,\,\,\, s \in \mathbb{R}\slash [\alpha, \beta] \\
  F(z)&=& \frac{1}{z} + \mathcal{O}(z^{-2}), \,\,\, z \to \infty.
\end{eqnarray*}
From this it is straightforward to deduce that
\begin{eqnarray}\label{CTidentity}
 2 F(z)  &=& V^\prime(z) - \frac{1}{2\pi i}R(z)h(z)
\end{eqnarray}

\section{Large N Expansions }

The fundamental theorem for establishing complete
large $N$ expansions of expectations of random variables related to
eigenvalue statistics  was developed in \cite{EM}.  A concise
statement of this result is:

\begin{thm}
\label{1PTASS}[Ercolani and McLaughlin, \cite{EM}]
There is $T>0$ and $\gamma > 0$ so that for all ${\bf t} \in
\mathbb{T}(T,\gamma)$, the following expansion holds true:
\begin{eqnarray}
\label{I.Weakasym}
\int_{-\infty}^{\infty} f(\lambda) \rho_{N}^{(1)}(\lambda) d \lambda =
f_{0} + N^{-2} f_{1} + N^{-4} f_{2} + \cdots,
\end{eqnarray}
provided the function $f(\lambda)$ is $C^{\infty}$ smooth, and grows no faster
than a polynomial for $\lambda \to \infty$.  The coefficients $f_{j}$
depend analytically on ${\bf t}$ for ${\bf t} \in \mathbb{T}(T, \gamma)$,
and the asymptotic expansion may be differentiated term by term.
\end{thm}

The complete details for the derivation of this result are presented
in \cite{EM};  however, there are a few specifics presented there that
we repeat here for use in subsequent sections and for general
background information:

\begin{itemize}
\item
The function $\rho_{N}^{(1)}$ has a full and uniform asymptotic
expansion, which starts off as follows:

\begin{eqnarray}
\rho_{N}^{(1)}(\lambda) = \psi(\lambda) + \mathcal{O} \left(
 N^{-1/2}\right).
\end{eqnarray}
\item
The specific form that this expansion takes depends very much on where
one is looking; for example, for $\lambda \in (\alpha, \beta)$, the
expansion takes:
\begin{eqnarray}
\label{3.002BUL}
& &
\rho_{N}^{(1)}(\lambda) = \psi(\lambda) + \frac{1}{4 \pi N} \left(
\frac{1}{\lambda - \beta} - \frac{1}{\lambda - \alpha}
\right) \cos{\left\{ N \int_{\lambda}^{\beta} \psi(s) ds  \right\} } \\
\nonumber
& &  \hspace{0.75in}+
\frac{1}{N^{2}} \left[ H(\lambda) + G(\lambda) \sin{\left\{N
\int_{\lambda}^{\beta} \psi(s) ds \right\} }\right] + \cdots
\end{eqnarray}
in which $H(\lambda)$ and $G(\lambda)$ are locally analytic functions
which are explicitly computable in terms of the original external field
$V(\lambda)$.

\end{itemize}

In \cite{EM} the primary application of this theorem was to establish
a complete large $N$  symptotic expansion of \ref{I.001} exists:

\begin{thm}\label{EQMSTHM}
There is $T>0$ and $\gamma > 0$ so that for $\bf{t} \in
\mathbb{T}(T,\gamma)$, one has the
$N \to \infty$ asymptotic
expansion
\begin{eqnarray}
\label{I.002} \ \ \ \log{
\left(\frac{Z_{N}(\bf{t})}{Z_{N}(\bf{0})} \right)} =
N^{2} e_{0}(x, {\bf{t}}) + e_{1}(x, {\bf{t}}) + \frac{1}{N^{2}}
e_{2}(x, {\bf{t}}) + \cdots.
\end{eqnarray}
The meaning of this expansion is:  if you keep terms up to order
$N^{-2h}$, the error term is bounded by $C N^{-2h-2}$, where the
constant $C$ is independent $\bf{t}$ for all $\bf{t}
\in \mathbb{T}(T,\gamma)$.
For each $j$, the function $e_{j}({\bf{t}})$ is an analytic
function of the (complex) vector $({\bf{t}})$, in a
neighborhood of $({\bf{0}})$. Moreover, the asymptotic
expansion of derivatives of $\log{ \left( Z_{N} \right)}$
may be calculated via term-by-term differentiation of the above
series. $\Box$
\end{thm}
\bigskip

{\bf Remark} Recently, Bleher and Its \cite{BI} have carried out a
similar asymptotic expansion of the partition function for a
1-parameter family of external fields.  A very interesting aspect
of their work is that they establish the nature of the asymptotic
expansion of the partition through a critical phase transition.
$\Box$

\bigskip
{\bf Remark} A subsequent application in \cite{EMP} is to develop a
hierarchy of
ordinary differential equations whose solutions determine recursively
the coefficients $e_{g}$ for potentials of the form $V(\lambda) =
\lambda^{2}/2 + t \lambda^{ 2 \nu}$.

\bigskip

{\bf Remark} The asymptotic results in \cite{EM} were also used
recently in \cite{Gustavsson} to establish that asympotics of each
individual eigenvalue have Guassian fluctuations, regardless of whether one is
in the bulk or near the edge of the spectrum (provided only that the
eigenvalue number, when counted from the edge, grows to $\infty$).

\bigskip

In the present paper we will make use of a mild extension of Theorem
\ref{1PTASS}, in which the function $f$ is of the form
$\frac{w(\lambda)
}{\lambda - z}$, with $z$ living outside of the inteval $[\alpha,
\beta]$:

\begin{thm}
\label{1PTASSNew}
For each $\delta > 0$, there is $T>0$ and $\gamma > 0$ so that for all
${\bf t} \in \mathbb{T}(T,\gamma)$, the following expansion holds true:
\begin{eqnarray}
\label{I.WeakNew}
\int_{-\infty}^{\infty} \left(\frac{w(\lambda)}{\lambda - z} \right) \rho_{N}^{(1)}(\lambda) d \lambda =
w_{0}(z) + N^{-2} w_{1}(z) + N^{-4} w_{2}(z) + \cdots,
\end{eqnarray}
provided $z \in \mathbb{C} \setminus [\alpha - \delta, \beta + \delta]
$, the function $w(\lambda)$ is analytic in a neighborhood of
$\mathbb{R}$, and grows no faster than a polynomial for $\lambda \to
\infty$.  The coefficients $w_{j}$
depend analytically on $z$ and ${\bf t}$ for ${\bf t} \in
\mathbb{T}(T, \gamma)$, and possess convergent Laurent expansions for
$z \to \infty$.  Furthermore, the asymptotic expansion may be
differentiated term by term.
\end{thm}
For $z$ bounded away from the real axis, this Theorem follows from
Theorem \ref{1PTASS}.  The mild extension to the case when $z$ may be
near the axis (but bounded away from the support $[\alpha, \beta]$)
follows by exploiting analyticity to replace the integral along the
real axis near $z$ by a semi-circular contour so that $\lambda$
remains uniformly bounded away from $z$.  Once the contour is such that
$\lambda$ is bounded away from $z$, the uniform asymptotic expansion
for $\rho_{N}^{(1)}$ may be used.  Since $z$ is away from the
support $[\alpha, \beta]$, the newly introduced semi-circular contour
is also bounded away from the support, and one may use arguments
similiar to those presented in \cite[Observation 4.2]{EM} (where they were
used for $\lambda$ real and outside the interval $[\alpha, \beta]$) to
show that $\rho_{N}^{(1)}(\lambda)$ is uniformly exponentially small on the
semi-circular contour, and also that the residue term obtained from
deforming the contour is also uniformly exponentially small.  We will
leave these details for the interested reader.

\section{Derivation of the Loop Equations}\label{loop}

We introduce some notation. Denote the Greens function of a random
matrix $M$ as
\begin{eqnarray*}
  G(z, M) &=& (z-M)^{-1}
\end{eqnarray*}
and and its trace as
\begin{eqnarray*}
  g(z) &=& \textrm{Tr}\, \, G(z).
\end{eqnarray*}
We evaluate $\partial G_{kl}/\partial M_{ij}$ in two different
ways to get a useful relation:

\begin{lem}
$$
\textbf{E}\left(G_{ki}G_{jl}\right) = N \textbf{E}\left(G_{kl}
V'(M)_{ji}\right)
$$
\end{lem}
\textbf{Proof.} Since $G\cdot (z-M)= \textbf{1},\,\,
\partial/\partial M_{ij}\,\, G\cdot (z-M) \equiv \textbf{0}$; or,
equivalently
\begin{eqnarray*}
  \partial G/\partial M_{ij}\,\, \cdot (z-M) - G \cdot E_{ij} &=&
  0,
\end{eqnarray*}
where $E_{ij}$ is the elementary permutation matrix with a $1$ in
the $ij^{th}$ entry and all other entries zero. It follows that
\begin{eqnarray*}
  \partial G/\partial M_{ij} &=& G \cdot E_{ij} \cdot G;
\end{eqnarray*}
in particular,
\begin{eqnarray*}
  \partial G_{kl}/\partial M_{ij} &=& \left(G_{ki}
  G_{jl}\right),
\end{eqnarray*}
and so
\begin{eqnarray*}
  \textbf{E}\left(\partial G_{kl}/\partial M_{ij}\right) &=&
  \textbf{E}\left(G_{ki}G_{jl}\right).
\end{eqnarray*}
On the other hand, integrating by parts yields
\begin{eqnarray*}
  \textbf{E}\left(\partial G_{kl}/\partial M_{ij}\right) &=&
  \frac{1}{\widetilde{Z}_N}\int_{\mathcal{H}}\frac{\partial G_{kl}}{\partial
  M_{ij}}  \exp\left\{-N \mbox{ Tr} [V_\textbf{t}(M)]\right\} dM \\
&=& -\frac{1}{\widetilde{Z}_N}\int_{\mathcal{H}} G_{kl}
\frac{\partial}{\partial
  M_{ij}} \exp\left\{-N \mbox{Tr} [V_\textbf{t}(M)]\right\} dM\\
&=& N \frac{1}{\widetilde{Z}_N}\int_{\mathcal{H}} G_{kl}
\mbox{Tr}\left(\nabla V(M)\cdot E_{ij}\right) \exp\left\{-N \mbox{Tr} [V_\textbf{t}(M)]\right\} dM\\
&=& N \frac{1}{\widetilde{Z}_N}\int_{\mathcal{H}} G_{kl}
V'(M)_{ji} \exp\left\{-N \mbox{Tr} [V_\textbf{t}(M)]\right\} dM\\
&=& N \textbf{E}\left(G_{kl} V'(M)_{ji}\right).
\end{eqnarray*}
Combining the above two representations for
$\textbf{E}\left(\partial G_{kl}/\partial M_{ij}\right)$ gives the
result. $\Box$
\begin{prop}
\begin{eqnarray} \label{quadreln}
\textbf{E}((g(z))^2) = N \textbf{E}\left(\mbox{Tr}\,\,(G \cdot
V'(M))\right)
\end{eqnarray}
\end{prop}
This follows directly from the lemma by setting $i=k$ and $j=l$,
summing over $k$ and $l$ and dividing by $1/N^2$ which yields
$$
\sum_{k,l}\textbf{E}\left(G_{kk}G_{ll}\right) =
N\sum_{k,l}\textbf{E}\left(G_{k,l}V'(M)_{lk}\right)
$$
\bigskip

The relation (\ref{quadreln}) can be naturally regarded as a
generating function for the second order matrix cumulants of $M$
when written in the equivalent form
\begin{eqnarray}\label{cum}
 \textbf{E}((g(z))^2) - \textbf{E}((g(z)))^2 &=& N
\textbf{E}\left(\frac{1}{N}\mbox{Tr}\,\,(G \cdot V'(M))\right) -
\textbf{E}((g(z)))^2.
\end{eqnarray}
To proceed further we will need to introduce some more notation.
First, we will use the following general expression for the
potential
$$
V(z) = \sum_{j=0}^\infty t_j z^j
$$
which is understood to have only finitely many $t_j$ non-zero. We
also have the formal \emph{vertex operator}
\begin{eqnarray}\label{vertex}
 \frac{d}{dV} &=& - \sum_{j=0}^\infty
 \frac{1}{z^{j+1}}\frac{d}{dt_j}.
\end{eqnarray}
(A precise  meaning for this formal relation will be given in the
beginning of  the next section.) This can be used to give a
compact formal representation of a generating function for matrix
moments in terms of the RM partition function (\ref{I.001}):
\begin{eqnarray} \label{moments}
  \frac{d}{dV} \frac{1}{N^2} \log Z_N &=& \sum_{j=0}^\infty
  \frac{1}{z^{j+1}} \textbf{E}\left(\frac{1}{N}\mbox{Tr}M^j\right).
\end{eqnarray}
\subsection{Asymptotic Expansions}
In order to make formal relations such as (\ref{moments})
meaningful we need to use some fundamental asymptotic facts. The
trace of $G(z)$ has a standard integral representation in terms of
the RM one-point function (\ref{one-point})
\begin{eqnarray}
  g(z) &=& N\int_{-\infty}^\infty
  \frac{\rho^{(1)}_N(\lambda)}{z-\lambda}d\lambda.
\end{eqnarray}
By boundedness and exponential decay of
$\rho^{(1)}_N(\lambda),
g(z)$ has a valid asymptotic expansion in
large $z$ as
\begin{eqnarray}
\int_{-\infty}^\infty
\frac{\rho^{(1)}_N(\lambda)}{z-\lambda}d\lambda  &\sim&
\sum_{j=0}^\infty \frac{1}{z^{j+1}}
\textbf{E}\left(\frac{1}{N}\mbox{Tr}M^j\right).
\end{eqnarray}
Thus (\ref{moments}) can be precisely understood as saying that
for each $m$ and large $z$,
\begin{eqnarray}\label{trunc}
\frac{d}{dV^{(m)}} \frac{1}{N^2} \log Z_N &=& \sum_{j=0}^{m-1}
\frac{1}{z^{j+1}} \int_{-\infty}^\infty
\frac{\lambda^j \rho^{(1)}_N(\lambda)}{z-\lambda}d\lambda \\
&=& \nonumber \int_{-\infty}^\infty
\frac{\rho^{(1)}_N(\lambda)}{z-\lambda}d\lambda +
\mathcal{O}(z^{-(m+1)})
\end{eqnarray}
where
$$
\frac{d}{dV^{(m)}} = - \sum_{j=0}^{m-1}
\frac{1}{z^{j+1}}\frac{d}{dt_j}.
$$
In a similar sense we have the following asymptotic equation for
each $m$ and large $z$:
\begin{eqnarray}\label{corr}
\nonumber \textbf{E}((g(z))^2) - \textbf{E}((g(z)))^2 &=&
\sum_{j,k= 0}^m
\frac{1}{z^{j+k+2}}\left(\textbf{E}\left(\mbox{TrM}^j\cdot
\mbox{TrM}^k\right) -
\textbf{E}\left(\mbox{Tr}M^j\right)\textbf{E}\left(\mbox{Tr}M^k\right)\right) + \mathcal{O}(z^{-(2m+3)})\\
&=& \nonumber \frac{d}{dV^{(m)}}\frac{d}{dV^{(m)}}\frac{1}{N^2} \log Z_N + \mathcal{O}(z^{-(2m+3)})\\
&=&  \frac{d}{dV^{(m)}} \int_{-\infty}^\infty
\frac{\rho^{(1)}_N(\lambda)}{z-\lambda}d\lambda +
 \mathcal{O}(z^{-(2m+3)}).
\end{eqnarray}
In what follows, we will use $\frac{d}{dV}$ instead of
$\frac{d}{dV^{(m)}}$ but with the above asymptotic interpretation
understood. In the rest of this section we need to establish that
there are estimates controlling the errors in the asymptotic
expansions (\ref{trunc}) and (\ref{corr}) that remain valid
uniformly as $N \to \infty$. To this end we first note that for
(\ref{trunc}) the error has the form
\begin{eqnarray*}
  \frac{1}{z^{m+1}}\int_{-\infty}^\infty \frac{\lambda^m \rho^{(1)}_N(\lambda)}{z-\lambda} d\lambda &=&
  \frac{1}{z^{m+1}}\left\{f^{(m)}_0(z) + N^{-2} f^{(m)}_1(z) + N^{-4} f^{(m)}_2(z) + \dots\right\}
\end{eqnarray*}
for $\textbf{t} \in \mathbb{T}(T,\gamma)$. The RHS is a uniformly
(in $\mathbb{T}$) valid asymptotic expansion which follows from
the fundamental Theorem \ref{1PTASS} . Similarly for (\ref{corr})
the error has the form
\begin{eqnarray*}
&-&\sum_{k=m}^\infty
\frac{1}{z^{k+1}}\frac{d}{dt_k}\frac{1}{z^{m+1}}\left\{f^{(m)}_0
(z)+ N^{-2} f^{(m)}_1(z) + N^{-4} f^{(m)}_2(z) + \dots\right\}\\
&=& \sum_{k=m}^\upsilon \frac{1}{z^{k+m+2}}\left\{g^{(m)}_0(z) +
N^{-2} g^{(m)}_1(z) + N^{-4} g^{(m)}_2(z) + \dots\right\}
\end{eqnarray*}
in which the sum on the RHS is finite since $V$ depends on only
finitely many distinct $t_k$. We use here the fact stated in
Theorem \ref{1PTASS} that these asymptotic expansions in $N$ can
be differentiated term by term, preserving uniformity.

With the observations of this section we may express the relation
(\ref{cum}) in terms of integral representations involving the
one-point function:
\begin{eqnarray} \label{vertmom}
 \frac{d}{dV}\int_{-\infty}^\infty
\frac{\rho^{(1)}_N(\lambda)}{z-\lambda}d\lambda &=& N^2
\int_{-\infty}^\infty
\frac{V'(\lambda)\rho^{(1)}_N(\lambda)}{z-\lambda}d\lambda - N^2
\left(\int_{-\infty}^\infty
\frac{\rho^{(1)}_N(\lambda)}{z-\lambda}d\lambda\right)^2,
\end{eqnarray}
to be understood in the sense of an asymptotic expansion in large
$z$ whose coefficients moreover have asymptotic expansions in
\textit{even} powers of $N$ which are uniform in admissible
\textbf{t}. We note that one consequence of this is that the two
terms on the RHS of (\ref{vertmom}) cancel at leading order so
that the difference has leading order $\mathcal{O}(1)$ for large
$N$.
\subsection{Loop Equations}
To prepare for the transformation to a recursive loop equation we
parse the first integral on the RHS of (\ref{vertmom})
as\begin{eqnarray} \label{chop}
  \int_{-\infty}^\infty
\frac{V'(\lambda)\rho^{(1)}_N(\lambda)}{z-\lambda}d\lambda &=&
\int_{\alpha - \delta}^{\beta + \delta}
\frac{V'(\lambda)\rho^{(1)}_N(\lambda)}{z-\lambda}d\lambda +
\mathcal{O} \left( e^{ - c N }  \right),
\end{eqnarray}
where $c > 0$ depends on the choice of the positive constant
$\delta$.  The justification for the exponential error term is
part of the proof of the fundamental asymptotic relation presented
in \cite{EM}.

From (\ref{chop}),we may further transform this term:
\begin{eqnarray} \nonumber
  \int_{-\infty}^\infty
\frac{V'(\lambda)\rho^{(1)}_N(\lambda)}{z-\lambda}d\lambda &=&
\frac{1}{2 \pi i} \int_{\alpha - \delta}^{\beta + \delta} \left(
\oint_{\mathcal{C}} \frac{V'(x)}{ x- \lambda}dx
\right)\frac{\rho^{(1)}_N(\lambda)}{z-\lambda}d\lambda +
\mathcal{O} \left( e^{ - c N }  \right),\\
\label{chop1} &=& \frac{1}{2 \pi i}
\oint_{\mathcal{C}}\int_{\alpha - \delta}^{\beta + \delta} \left(
\frac{V'(x)}{ x- \lambda}
\right)\frac{\rho^{(1)}_N(\lambda)}{z-\lambda}d\lambda \ dx+
\mathcal{O} \left( e^{ - c N }  \right),\\
\label{chop2} &=& \frac{1}{2 \pi i}
\oint_{\mathcal{C}}\int_{\alpha - \delta}^{\beta + \delta}  V'(x)
\rho^{(1)}_N(\lambda) \left( \frac{1}{x-\lambda} - \frac{1}{z -
\lambda} \right) \left(\frac{1}{z-x} \right) d\lambda \ dx+
\mathcal{O} \left( e^{
- c N } \right)\\
\label{chop3} &=& \frac{1}{2 \pi i}
\oint_{\mathcal{C}}\int_{\alpha - \delta}^{\beta + \delta}  V'(x)
\rho^{(1)}_N(\lambda) \left( \frac{1}{x-\lambda}\right)
\left(\frac{1}{z-x} \right) d\lambda \ dx+ \mathcal{O} \left( e^{
- c N } \right)
\\
\label{chop4} &=& \frac{1}{2 \pi i}
\oint_{\mathcal{C}}\frac{V'(x)}{z-x}\int_{\alpha - \delta}^{\beta
+ \delta} \frac{ \rho^{(1)}_N(\lambda)}{ x-\lambda} d\lambda \ dx+
\mathcal{O} \left( e^{ - c N } \right),
\end{eqnarray}
where in the first line we have expressed $V'(\lambda)$ as a loop
integral a la Cauchy's Theorem, with the contour $\mathcal{C}$
encircles the interval $(\alpha - \delta, \beta + \delta)$, with
$z$ outside the contour of integration, and in (\ref{chop3}), one
term has vanished by Cauchy's Theorem and analyticity.

Inserting (\ref{chop4}) into (\ref{vertmom}), we have derived the
final form of the Loop Equation generating function.

\begin{eqnarray} \label{LoopEQNGF}
\ \ \ \ \ \ \ \frac{1}{2 \pi i}
\oint_{\mathcal{C}}\frac{V'(x)}{z-x}\int_{\alpha - \delta}^{\beta
+ \delta} \frac{ \rho^{(1)}_N(\lambda)}{ x-\lambda} d\lambda \ dx
&=& - N^{-2} \frac{d}{dV}\int_{\alpha - \delta}^{\beta + \delta}
\frac{\rho^{(1)}_N(\lambda)}{z-\lambda}d\lambda  -
\left(\int_{\alpha - \delta}^{\beta + \delta}
\frac{\rho^{(1)}_N(\lambda)}{z-\lambda}d\lambda\right)^2 +
\mathcal{O} \left( e^{ - c N } \right).
\end{eqnarray}

Using Theorem \ref{1PTASSNew}, the term $\int_{\alpha
-\delta}^{\beta +
\delta}\frac{\rho^{(1)}_N(\lambda)}{z-\lambda}d\lambda$ is easily
seen to possess an asymptotic expansion in large $N$, each of
whose coefficients possesses a Laurent expansion in large $z$:
\begin{eqnarray} \label{Pgexp}
\int_{\alpha - \delta}^{\beta + \delta}
\frac{\rho^{(1)}_N(\lambda)}{z-\lambda}d\lambda \sim
\sum_{g=0}^{\infty}N ^{- 2 g} \ P_{g}(z).
\end{eqnarray}
Combining (\ref{trunc}) and (\ref{I.002}) we see that
\begin{eqnarray}\label{enum}
 P_{g}(z) &=& \frac{d}{dV} e_g(\textbf{t}) = \frac{d}{dV^{(\upsilon + 1)}}
 e_g(\textbf{t}) = - \sum_{j=0}^{\upsilon}
\frac{1}{z^{j+1}}\frac{d e_g(\textbf{t})}{dt_j}.
\end{eqnarray}
 Inserting
(\ref{Pgexp}) into the Loop Equation generating function
(\ref{LoopEQNGF}) yields the hierarchy of Loop Equations:

\begin{eqnarray} \label{LoopEQNFIN}
\ \ \ \ \ \ \ \frac{1}{2 \pi i}
\oint_{\mathcal{C}}\frac{V'(x)}{z-x} P_g(x)dx &=& -
\frac{d}{dV} P_{g-1}(z)  - \sum_{g'=0}^g P_{g'}(z)P_{g-g'}(z)\\
\ \ \ \ \ \ \ \frac{1}{2 \pi i}
\oint_{\mathcal{C}}\frac{\left(V'(x) - 2P_0(x)\right)}{z-x} P_g(x)
dx &=& \nonumber -  \frac{d}{dV} P_{g-1}(z)  - \sum_{g'=1}^{g-1}
P_{g'}(z)P_{g-g'}(z)\\
\ \ \ \ \ \ \ \frac{1}{2 \pi i}
\oint_{\mathcal{C}}\frac{\psi(x)}{x-z} P_g(x) dx &=& \nonumber
\frac{d}{dV} P_{g-1}(z)  + \sum_{g'=1}^{g-1} P_{g'}(z)P_{g-g'}(z)
\end{eqnarray}
where the transition to the final recursion formula is mediated by
the identity (\ref{CTidentity})
\begin{eqnarray}\label{CTidentity2}
\psi(x) = V'(x) - 2\int_{-\infty}^\infty
\frac{\psi(\lambda)}{x-\lambda}d\lambda = V'(x) - 2P_0(x)
\end{eqnarray}
where $\psi(x)$ here is interpreted as the analytic extension of
the density for the equilibrium measure off of the slit $[\alpha,
\beta]$ as given by (\ref{psi}).

\bigskip

With this result in hand it is now possible to consider a
recursive derivation of the terms $P_g$ starting with $P_0$ as
given by (\ref{CTidentity2}). These terms are related to the map
enumeration generating functions, $e_g(\textbf{t})$, through
(\ref{enum}). In the physics literature there are instances in
which loop equations are used to formally derive expressions for
some of the $e_g$. In particular, we refer the reader to
\cite{Ambjorn}.

A natural application of our derivation of (\ref{LoopEQNFIN})
would be to the derivation of closed form expressions for
$e_g(\textbf{t})$ which extend our results in \cite{EMP} for
potentials $V$ depending only on a single non-zero time
$t_{2\nu}$. We may also be able to use these equations to say
something about the general qualitative and asymptotic behavior of
the $e_g$. Finally, the $P_g(z,\textbf{t})$ contain information
about the large $N$ asymptotic behavior of the matrix moments,
such as (\ref{matmoments}), which could be used to explore whether
or not the general unitary ensembles are asymptotically free.

\bibliographystyle{amsplain}

\end{document}